\documentclass[conference]{IEEEtran}

\usepackage{cite}
\usepackage{amsmath,amssymb,amsfonts}
\usepackage{textcomp}

\usepackage{graphicx}

\usepackage{cite}
\usepackage{amsmath}
\usepackage{amssymb}
\usepackage{amsfonts}
\usepackage[hidelinks]{hyperref}
\usepackage{xcolor}

\usepackage{algorithm}
\usepackage[noend]{algpseudocode}

\usepackage{amsthm}
\usepackage{caption}
\usepackage{subcaption}

\newcommand*\Input[1]{\Statex \textbf{Input:} #1}
\newcommand*\Output[1]{\Statex \textbf{Output:} #1}
\algrenewcommand\alglinenumber[1]{#1}
\algrenewcommand\algorithmicindent{1em}

\usepackage{caption}
\usepackage{subcaption}

\newtheoremstyle{def_style}
  {}          
  {}          
  {}          
  {}          
  {\bfseries} 
  {.}         
  {.5em}      
  {}          

\theoremstyle{def_style}
\newtheorem{define}{Definition}

\theoremstyle{def_style}
\newtheorem{lemma_}{Lemma}

\theoremstyle{def_style}

\DeclareMathOperator*{\argmax}{arg\,max}

\usepackage{enumitem}
\setlist[itemize]{leftmargin=*, topsep=0pt, itemsep=0pt, parsep=0pt, partopsep=0pt}
\setlist[enumerate]{leftmargin=*, topsep=0pt, itemsep=0pt, parsep=0pt, partopsep=0pt}

\begin{document}

\title{Sketching Multidimensional Time Series\\for Fast Discord Mining}

\author{\IEEEauthorblockN{Chin-Chia Michael Yeh, Yan Zheng, Menghai Pan, Huiyuan Chen, Zhongfang Zhuang,\\Junpeng Wang, Liang Wang, Wei Zhang, Jeff M. Phillips$^\dagger$, Eamonn Keogh$^\ddagger$}
\IEEEauthorblockA{\textit{Visa Research}, \textit{University of Utah}$^\dagger$, \textit{University of California, Riverside}$^\ddagger$ \\
\{miyeh,yazheng,menpan,hchen,zzhuang,junpenwa,liawang,wzhan\}@visa.com}
}

\maketitle              

\begin{abstract}
Time series discords are a useful primitive for time series anomaly detection, and the matrix profile is capable of capturing discord effectively.
There exist many research efforts to improve the scalability of discord discovery with respect to the \textit{length} of time series. 
However, there is surprisingly little work focused on reducing the time complexity of matrix profile computation associated with \textit{dimensionality} of a multidimensional time series.
In this work, we propose a sketch for discord mining among multi-dimensional time series. 
After an initial pre-processing of the sketch as fast as reading the data, the discord mining has runtime independent of the dimensionality of the original data.  
On several real world examples from water treatment and transportation, the proposed algorithm improves the throughput by at least an order of magnitude (50X) and only has minimal impact on the quality of the approximated solution. 
Additionally, the proposed method can handle the dynamic addition or deletion of dimensions inconsequential overhead. 
This allows a data analyst to consider ``what-if" scenarios in real time while exploring the data. 
\end{abstract}



\section{Introduction}
Time series discords are a simple, effective, and robust primitive for detecting anomalies in time series data~\cite{keogh2005hot,chandola2009anomaly,wu2021current}.
While there have been dozens of algorithms proposed to compute discords in the last twenty years, in recent years the \textit{matrix profile} (MP) has emerged as the most effective and versatile computation tool for discovering discords~\cite{yeh2016matrix,yeh2018time}.
There have been numerous efforts on improving the scalability of the MP. 
For example, Zimmerman et al.~\cite{zimmerman2018scaling} improve the computational speed by exploiting hardware, Zhu et al.~\cite{zhu2018matrix} introduced an anytime algorithm with a fast convergence rate, and efficient approximation of the MP is proposed in~\cite{zimmerman2019matrix}.
However, all these ideas only consider univariate time series.  
In other words, these works focus on reducing the time complexity with respect to the \textit{length} of the input time series, while the runtime cost that is associated with \textit{dimensionality} of multidimensional time series is unchecked.
The word ``\textit{dimensionality}" is used inconsistently in the time series literature. 
To be clear, in this work \textit{dimensionality} means the number of individual and concurrent streams, for example, a three-dimensional medical time series might contain \texttt{\{ECG|respiration|temperature\}}.  

The ability to scale with respect to dimensionality is crucial in many domains.
For instance, there are hundreds of millions of merchants in modern-day financial transaction networks.
If a regulatory agency wants to monitor the activity of each merchant for detecting suspicious behavior in real-time, it will need to maintain hundreds of millions of MPs for discord discovery. 
In this paper, we propose a sketching algorithm on multidimensional time series for discord mining so that it can dramatically reduce the cost associated with monitoring multiple dimensions simultaneously.

\begin{figure}[htbp]
\centerline{
\includegraphics[width=0.95\linewidth]{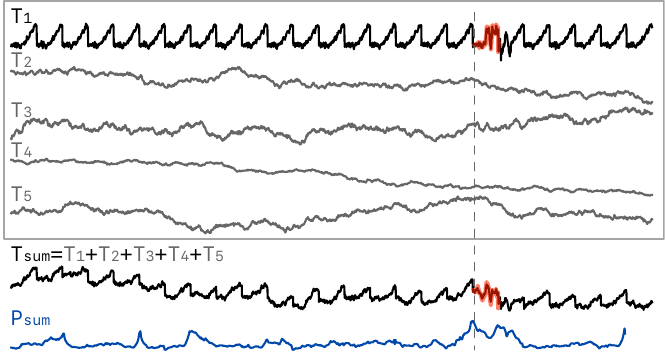}
}
\caption{
In the multidimensional time series $\mathbf{T}=[T_1, ..., T_5]$, the time series discord occurs in $T_1$ can be discovered even if the matrix profile~$P_{sum}$ is computed using the summed time series $T_{sum}=T_1+\cdots+T_5$.
}
\label{fig:toy}
\end{figure}

To preview our ideas, consider the example shown in Fig.~\ref{fig:toy} which demonstrates how simple addition-based dimension reduction works for multidimensional time series discord mining. 
The multidimensional time series $\mathbf{T}$ has five dimensions $[T_1,...,T_5]$, and the time series discord occurs in $T_1$.
We can add all the dimensions together (i.e., $T_{sum}$) and compute the associated MP (i.e., $P_{sum}$). 
Due to the robustness of the MP method, $P_{sum}$ still reveals the temporal location of the discord that occurred only in $T_1$ despite the existence of the other irrelevant dimensions.
Because we only need to compute one MP instead of five MPs, a 5X improvement in speed is achieved.
Before continuing, we wish to ward off a potential misunderstanding. 
Clearly, it is not meaningful to add dollars and yen, much less dollars and temperature. 
However, we \textit{can} meaningfully add $z$-normalized time series of arbitrary origin. 
The $z$-normalization step converts the data into unitless shapes.

As pointed out in~\cite{yeh2017matrix}, if more and more irrelevant dimensions are included, eventually the discord in $T_1$ will be missed if only the MP of the summed time series is examined.
To solve this problem, we designed a sketch~\cite{charikar2002finding,clarkson2017low,yeh2022embedding} where MP can be built on the sketch of the multidimensional time series and discord property can be preserved.
The contributions of this paper include:
\begin{itemize}
    \item This is the first work to apply sketching techniques to multidimensional time series discord mining with theoretical guarantees.  
    \item After computing the sketch in time as fast as reading the data, we guarantee to find time points of significant discords with high probability efficiently and in time independent of the number of dimensions.   
    \item Experiments demonstrate that with an insignificant drop in accuracy, the sketched discord mining can achieve a 50X speed-up.   This utility is realized on several real-world data sets from water treatment, distributed systems, and transportation.  
\end{itemize}
\section{Background}

\subsection{Related Work}
Our proposed ideas have an interesting historic context, in that they resemble \textit{Dorfman} testing (\textit{group} testing, \textit{pooled} testing, etc.)~\cite{dorfman1943detection}. 
This is a procedure used to reduce the cost of screening a large number of individuals for infectious diseases. 
It works by compositing a set of individual specimens (e.g., blood or urine) into a common pool. 
If the pool tests negative, all individuals within it are diagnosed as negative. 
If the pool tests positive, retesting is needed to find the positive individual(s). 
The reader will appreciate that in our example in Fig.~\ref{fig:toy}, we had a ``pool" of five time series, and the summed time series did indeed ``test positive" for a discord. 

How to efficiently pool data into a compressed representation while provably preserving critical information has taken on the name of \emph{sketching}.  These ideas have proven invaluable in domains such as network monitoring~\cite{charikar2002finding,cormode2005improved}, linear algebra~\cite{woodruff2014sketching,ghashami2016frequent}, machine learning~\cite{clarkson2017low,munteanu2018coresets,karnin2019discrepancy}, and spatial statistics~\cite{matheny2016scalable}.  
To the best of our knowledge, this is the first time that \textit{Dorfman testing} or \emph{sketching} has been used in time series analysis.

In recent years, there has been an increasing focus on multidimensional time series~\cite{minnen2007detecting,yeh2017matrix,li2019mad,gao2019discovering,shih2019temporal,yeh2021online}.
For example, Yeh et al.~\cite{yeh2017matrix} adopted the matrix profile~\cite{yeh2016matrix} idea for mining motifs in multidimensional time series.
Gao et al.~\cite{gao2019discovering} developed a variable-length subdimensional motif discovery system for multidimensional time series.
Similar to our work, the system introduced in~\cite{minnen2007detecting} also used a type of random projection as a subroutine.
However, instead of using it for mining time series discords, the proposed system in~\cite{minnen2007detecting} is a subdimensional motif discovery system similar to~\cite{gao2019discovering}.
Other tasks, such as time series forecasting~\cite{shih2019temporal,yeh2021online} and anomaly detection~\cite{li2019mad} have also been studied for multidimensional time series.
Among these works, research efforts on time series anomaly detection are more relevant to our work because time series discords not only can be utilized to solve the anomaly detection problem but many independent papers have shown they are very competitive compared to the state-of-the-art~\cite{anton2018time, wu2021current,daigavane2020detection,nakamura2020merlin}.
However, none of these prior works provide a scalable discord mining algorithm for multidimensional time series.

The matrix profile is an efficient way to discover discords in time series~\cite{yeh2016matrix}.
In recent years, there have been many attempts to further improve the computational speed of the matrix profile algorithm through approximation~\cite{zhu2018matrix,zimmerman2019matrix,yeh2022error}.
Nevertheless, all of the aforementioned works focused on reducing the computational cost associated with the \textit{length} of the input time series rather than the \textit{dimensionality} of the time series.
These works cannot be used to resolve the scalability problem raised in this paper.
Dimensionality reduction methods such as PAA~\cite{keogh2001dimensionality}, DWT~\cite{chan1999efficient}, and DFT~\cite{faloutsos1994fast} are widely used in time series data mining to reduce run time~\cite{wang2013experimental}.
However, as we noted earlier, here dimensionality refers to the \textit{length} of the time series, not the number of time series; therefore, they also do not address the scalability problem associated with dimensionality.

\subsection{Definitions and Notation}
We begin by defining the data type of interest, \textit{time series} and \textit{multidimensional time series}:

\begin{define}
    A \textit{time series} $T \in \mathbb{R}^{n}$ is an array of real valued numbers $t_i \in \mathbb{R}:T=\left[t_1, t_2, ..., t_n \right]$ where $n$ is the length of $T$.
    A \textit{multidimensional time series} $\mathbf{T} \in \mathbb{R}^{d \times n}$ is a collection of single dimensional time series $\mathbf{T}=\left[T^{(1)}, T^{(2)}, ..., T^{(d)} \right]$ where $d$ is the dimensionality of $\mathbf{T}$.
\end{define}

Since time series discords are a \textit{local} property of a time series, we are not interested in the \textit{global} properties of a time series, but the \textit{local} \textit{subsequences}~\cite{yeh2018time}:

\begin{define}
    A \textit{subsequence} $T_{i,m} \in \mathbb{R}^{m}$ of a time series~$T$ is a length~$m$ contiguous subarray of $T$ starting from position~$i$. 
    Formally, $T_{i,m}=[t_{i}, t_{i+1}, ..., t_{i+m-1}]$.
    Note, we use $T_{i,m}^{(j)}$ to denote the $i$th subsequence in $j$th dimension of a multidimensional time series~$\mathbf{T}$.
\end{define}


Because time series discords concern the nearest neighbor (\textit{1NN}) relation between a query subsequence and all subsequences of a given time series, we define a \textit{1NN\_dist} function which computes the nearest neighbor distance between the query subsequence with a given time series.

\begin{define}
\sloppy
    Given a query subsequence~$T_{q,m} \in \mathbb{R}^m$ and a time series $T \in \mathbb{R}^n$, a \textit{1NN distance function}~$\textit{1NN\_dist}(T_{q,m}, T)$ computes and returns the distance between $T_{q,m}$ and the nearest neighbor subsequence in~$T$, i.e., $\min_{T_{i,m} \in T} \textit{dist}(T_{q,m}, T_{i,m})$ where $\textit{dist}(\cdot, \cdot)$ computes the $z$-normalized Euclidean distance between the two inputs.
\end{define}

\sloppy
We use $z$-normalized Euclidean distance function in this work.
Note, $\textit{1NN\_dist}(\cdot, \cdot)$ and $\textit{dist}(\cdot, \cdot)$ look similar, but they are very different functions.
$\textit{1NN\_dist}()$ computes the one nearest neighbor distance between a subsequence and a time series and $\textit{dist}()$ computes the distance (e.g., $z$-normalized Euclidean distance) between two subsequences.
The particular local properties that we seek to capture are \textit{time series discords}:



\begin{define}
\label{def:discord_2ts}
    Given a training time series~$T_{\text{train}}$, a testing time series~$T_{\text{test}}$ and a subsequence length~$m$, the \textit{time series discord} of~$T_{\text{test}}$ is the subsequence of length~$m$ in~$T_{\text{test}}$ with largest nearest neighbor distance to subsequences in~$T_{\text{train}}$, i.e., $\argmax_{T_{i,m} \in T_{\text{test}}} \textit{1NN\_dist}(T_{i,m}, T_{\text{train}})$. When detecting the discord within a single time series, i.e. $T_{\text{test}}$ = $T_{\text{train}}$, the definition is also applied.
\end{define}

Additionally, since we are considering multidimensional time series, we extend the aforementioned single-dimensional time series discord to the following:

\begin{define}
\label{def:multi_discord}
    Given a multidimensional training time series~$\mathbf{T}_{\textrm{train}}=\left[T^{(1)}_{\text{train}}, T^{(2)}_{\text{train}}, ..., T^{(d)}_{\text{train}} \right]$, a multidimensional testing time series~$\mathbf{T}_{\textrm{test}}=\left[T^{(1)}_{\text{test}}, T^{(2)}_{\text{test}}, ..., T^{(d)}_{\text{test}} \right]$ and a subsequence length~$m$, the \textit{time series discord} of~$\mathbf{T}_{\textrm{test}}$ is the subsequence of length~$m$ in~$\mathbf{T}_{\textrm{test}}$ with the largest nearest neighbor distance (where the nearest neighbors are in~$\mathbf{T}_{\text{train}}$ of the same dimension as the subsequence).
    Formally:
    \begin{equation*}
    \footnotesize
        T^{(j^*)}_{i^*,m} = 
        {\argmax_{\substack{T_{i,m}^{(j)} \in T_{\text{test}}^{(j)} \\ T_{\text{test}}^{(j)} \in \mathbf{T}_{\text{test}}}} 
        \textit{1NN\_dist}(T_{i,m}^{(j)}, T_{\text{train}}^{(j)})}
    \end{equation*}
The basic time series discord only needs to return the time index $i^*$ at the start of the subsequence, $T^{(j^*)}_{i^*,m}$.  The \emph{dimensional time series discord} identifies both the time index $i^*$, and also the dimension $j^*$ in which this most anomalous pattern occurs.  
\end{define}

The most efficient method of locating the single dimension time series discord exactly is the \textit{matrix profile}~\cite{yeh2016matrix,yeh2018time}.



\begin{define}
\label{def:mp_2ts}
A ab-join matrix profile~\cite{yeh2018time}~$P \in \mathbb{R}^{n_{\text{test}} - m + 1}$ of a time series~$T_{\text{test}} \in \mathbb{R}^{n_{\text{test}}}$ and a time series~$T_{\text{train}} \in \mathbb{R}^{n_{\text{train}}}$ is a meta time series annotating $T_{\text{test}}$ that stores the distance between each subsequence of length~$m$ in $T_{\text{test}}$ and its nearest neighbor in $T_{\text{train}}$. When $T_{\text{test}} = T_{\text{train}}$, it is the definition of \textit{self-join} matrix profile~\cite{yeh2018time}.
\end{define}
The matrix profile~$P$ of a time series~$T$ ($|T| = n$) is shown in Fig.~\ref{fig:mp}.
The length of~$P$ is $n - m + 1$ where $m$ is the subsequence length.
By examining the location of the largest value in~$P$, the embedded time series discord can be discovered.

\definecolor{discord}{HTML}{FF0000}

\begin{figure}[htbp]
\centerline{
\includegraphics[width=0.95\linewidth]{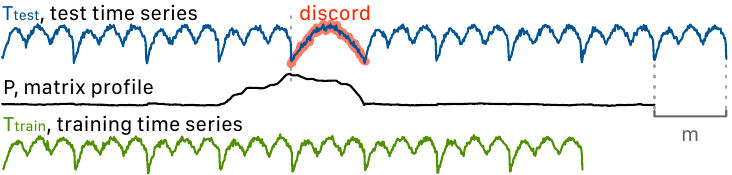}
}
\caption{
The matrix profile~$P$ of the time series $T_{\text{test}}$ is obtained by joining it with another time series $T_{\text{train}}$.
The highest point on~$P$ correspond to the locations of discord (\textcolor{discord}{red}). 
}
\label{fig:mp}
\end{figure}

The matrix profile can be computed exactly in $O(n_{\text{train}} n_{\text{test}})$ time complexity.
Since the multidimensional time series discord is defined as the single dimension discord with largest nearest neighbor distance (Def.~\ref{def:multi_discord}), the multidimensional time series discord can be discovered exactly by running matrix profile $d$ times for a $d$ dimensional time series.
In other words, the time complexity for this basic solution is $O(d \cdot n_{\text{train}}n_{\text{test}})$, and the goal of this work is to further reduce this complexity.

\section{Algorithm}
The proposed method has two stages: 1) the sketching stage and 2) the detection stage.
In the sketching stage, each dimension is randomly assigned to different groups.
With the group assignment initialized, the sketched time series are used in the second stage for fast detection of time series discords.
In addition to introducing the two stages of the proposed method, we also discuss how the proposed method handles the addition/deletion of dimensions.

\subsection{Sketching}
The sketching method is based on a count sketch~\cite{charikar2002finding}.  It maps from $d$ time series (one for each dimension) to $k$ time series, one for each group.  
The count sketch uses two pair-wise independent hash functions $h \in \mathcal{H}$ with $h : [d] \to [k]$ and $s \in \mathcal{S}$ with $s : [d] \to \{-1,+1\}$.  
We use $[d]$ as a short hand for $\{0, 1, \ldots, d-1\}$.  
The first hash function $h$ maps each dimension to a group, the second one endows each dimension with a sign.  
Recall that the hash functions $h$ and $s$ are deterministic, but their selection from the families $\mathcal{H}$ and $\mathcal{S}$ is random.    

For a multidimensional time series $\mathbf{T}=\left[T^{(1)}, T^{(2)}, ..., T^{(d)} \right]$ 
 and each group index $g \in [k]$, let $J_g = \{j \in [d] \mid h(j) = g\}$ be the set of dimensions mapped to group $g$.  Then these encode an aggregated time series $\textbf{R}^{(g)} = \sum_{j \in J_g} s(j) {T}^{(j)}$.  

As is common in sketching, the algorithm (see Alg.~\ref{alg:projection}) is simple, and in this case takes $O(nd)$ time, basically just reading the input data.  
This can be done to create $\textbf{R}_{\textrm{train}}\leftarrow \textsc{Sketch}(\bar{\textbf{T}}_{\textrm{train}}, k)$ and  $\textbf{R}_{\textrm{test}} \leftarrow \textsc{Sketch}(\bar {\textbf{T}}_{\textrm{test}}, k)$ using the same hashing functions, where $\bar{\textbf{T}}_{\textrm{train}}$ and $\bar {\textbf{T}}_{\textrm{test}}$ are the z-normalized multidimensional time series of the train and test data, respectively.




\begin{algorithm}[ht]
\footnotesize
    \centering
    \caption{
    Sketching Algorithm\label{alg:projection}}
    \begin{algorithmic}[1]
        \Input{multidimensional time series $\mathbf{T} \in \mathbb{R}^{d \times n}$, sketch dimension~$k$}
        \Output{sketched matrix~$\mathbf{R} \in \mathbb{R}^{k \times n}$}
        \Function{Sketch}{$\mathbf{T}, k$}
          \State $\mathbf{R} \gets $ zero matrix with size $k \times n$
          \For{$j \in [0, \cdots, d-1]$}
            \State $i = h(j)$
            \State $\mathbf{R}^{(i)} = \mathbf{R}^{(i)} + s(j) {T}^{(j)}$
          \EndFor
          \State Return $\mathbf{R}$
        \EndFunction    
    \end{algorithmic}
\end{algorithm}

\subsection{Detection}
Given the sketch matrices $\mathbf{R}_{\text{train}}$ and $\mathbf{R}_{\text{test}}$ for the training and testing multidimensional time series, and the subsequence length~$m$, we can use Alg.~\ref{alg:detection} to find the top-1 time series discord.
The function $\textsc{ABjoinMP}(\cdot)$ returns both the index and the score for the discord.

\begin{algorithm}[ht]
\footnotesize
    \centering
    \caption{
    Discord Time Detection Algorithm\label{alg:detection}}
    \begin{algorithmic}[1]
        \Input{sketched training time series~$\mathbf{R}_{\text{train}}$, testing time series~$\mathbf{R}_{\text{test}}$, subsequence length~$m$}
        \Output{discord time index~$i^*$, discord group~$g^*$, }
        \Function{Time-Detection}{$\mathbf{R}_{\text{train}}$, $\mathbf{R}_{\text{test}}$, $m$}
        \State $\hat{s}_{\text{bsf}} \gets 0$  
        \For{$g \in [0, \cdots, k - 1]$}
          \State $i, s^{(g)} \gets \textsc{ABjoinMP}(R_{\text{train}}^{(g)}, R_{\text{test}}^{(g)}, m)$ \Comment{Def.~\ref{def:mp_2ts}}
          \If{$s^{(g)} > \hat{s}_{\text{bsf}}$}
            \State $g_{\text{bsf}} \gets g$, $\hat{s}_{\text{bsf}} \gets \hat{s}^{(g)}$,  $i_{\text{bsf}} \gets i$
          \EndIf
        \EndFor
        \State \Return $i^* \gets i_{\text{bsf}}$, $g^* \gets g_{\text{bsf}}$
        \EndFunction
    \end{algorithmic}
\end{algorithm}

\begin{algorithm}[ht]
\footnotesize
    \centering
    \caption{
    Discord Dimension Detection Algorithm\label{alg:dim-detection}}
    \begin{algorithmic}[1]
        \Input{training time series~$\mathbf{T}_{\text{train}}$, testing time series subsequences~$\mathbf{T}_m$, subsequence length~$m$, group of time series $J_g$}
        \Output{discord dimension index~$j^*$}
        \Function{Dimension-Detection}{$\mathbf{T}_{\text{train}}$, $\mathbf{T}_m$, $m$, $J_{g^*}$}
        \State $s_{\text{bsf}} \gets 0$ 
        \For{$j \in J_{g^*}$}
          \State $\_, s^{(j)} \gets \textsc{ABjoinMP}(T_{\text{train}}^{(j)}, T^{(j)}_m, m)$ \Comment{Def.~\ref{def:mp_2ts}}
          \If{$s^{(j)} > s_{\text{bsf}}$}
            \State $j^* \gets j$, $s_{\text{bsf}} \gets s^{(j)}$
          \EndIf
        \EndFor
        \State \Return $j^*$
        \EndFunction
    \end{algorithmic}
\end{algorithm}

The discord detection algorithm runs in two phases.  
In the first phase to detect the time of the discord, Alg.~\ref{alg:detection}, it only considers the sketched time series $\mathbf{R}_{\text{train}}$ and $\mathbf{R}_{\text{test}}$, and identifies the time of the discord subsequence in the test data $i^*$, and the group $g^*$ of dimensions which may contain the discord subsequence.  
To do this, it treats the sketched time series as just $k$-dimensional time series, and checks each of the $k$ possible sketched-to dimensions.
Since this phase only operates on the sketched data, its runtime is independent of the number of dimensions $d$.  

In the second phase showed in Alg.~\ref{alg:dim-detection}, it recovers the dimension which leads to the discord.
It takes in the test data subsequences $\textbf{T}_m$ of multidimensional time series, starting at index $i$ for a length of $m$ that contains all dimensions from the test set for that time window; it actually only needs those for indexes $j \in J_{g^*}$ in group $g^*$.  
Then for this fixed time window in the test data, it checks each of the dimensions that contributed to group $g^*$ to see which one is the discord using standard Matrix Profile ab-join.
Note that users have the option to refine the current approximated solution by performing an additional Matrix Profile join operation on the entire sequence of the identified dimension to find a subsequence with an even higher discord score. 
This functionality has been implemented in the released source code.

We use the matrix profile~\cite{yeh2018time} method to locate and score the time series discord within a given pair of training/testing single-dimension time series for the \textsc{Time-Detection} algorithm.
In particular, we use the SCAMP algorithm~\cite{zimmerman2018scaling} to compute the matrix profile.
Because SCAMP is not the only matrix profile algorithm, it is possible to replace it with a faster and approximated algorithm from~\cite{zhu2018matrix,zimmerman2019matrix,yeh2022error}.

\sloppy
The time complexity of the discord time detection is $O(k \cdot n_{\text{train}} n_{\text{test}})$, where the $n_{\text{train}} n_{\text{test}}$ part of the big-O notation is from the matrix profile algorithm (see Def.~\ref{def:mp_2ts}).
To detect the dimension of the discord, another $O((d/k) n_{\text{train}})$ time is needed.  
The exact multidimensional discord mining algorithm, without the $O(d (n_{\text{train}} + n_{\text{test}}))$ sketching step is $O(d \cdot n_{\text{train}} n_{\text{test}})$.


To extend the proposed method to the single time series scenario, the user can input the same multidimensional time series to both $\textbf{T}_{\text{train}}$ and $\textbf{T}_{\text{test}}$; then use the self-join variate of the matrix profile algorithm instead of the ab-join variant in Alg.~\ref{alg:detection}.
It is also possible to extend the method to streamline time series.
To achieve such extension, lines 4-5 in Alg.~\ref{alg:projection} need to be implemented to only invoke whenever a new test data is received.
The matrix profile algorithm used in Alg.~\ref{alg:detection} will also need to be replaced with a streaming variant such as the ones presented in~\cite{yeh2018time,zimmerman2019matrix,yeh2022error}.

\subsection{Addition/Deletion of Dimensions}
The proposed method can also be extended to handle the addition or deletion of dimensions.  Since the count sketch is ``linear'' it can be updated, via additions, deletions, and modifications.  
Such scenarios could happen when sensors are added or removed from a system.
For instance, in the case where dimension $j \in [d]$ is deleted, the $g$th sketched time series where $g = h(j)$ is updated as 
$\textbf{R}^{(g)} = \textbf{R}^{(g)} - s(j) \textbf{T}^{(j)}$.  If only the $i$th time point of the $j$th dimension is updated, to increase its value by a value $\delta$, then we can update $R^{(g)}_i = R^{(g)}_i + s(j) \delta$.   
The detection algorithms are unchanged.  

As a technical issue, it may not be appropriate to add a new time series dimension that is only available for part of the training time.  At the beginning of the membership, it could create an artificial ``step" shape.  While we may be able to mark this start point as to be avoided in the Discord Time Detection algorithm, we mostly recommend avoiding this complication, and suggest that 
only time series that starts at the same time should be added together.

\subsection{Analysis of Accuracy of Algorithm}
We analyze the accuracy of the algorithm by extending the analysis of the CountSketch~\cite{charikar2002finding}.  
In particular, we analyze the probability that a discord subsequence of length $m$ is detected as the discord in the sketched representation.  
The details of the analysis are in Appendix.  We show several results.  

First, if a time series $T^{(j)}$ is randomly placed in group $g$, and multiplied by sign $s(j) \in \{-1,+1\}$ in the sketched time series $\textbf{R}^{(g)}$, then for any point $i$ the expected value $E[s(j) \cdot R^{(g)}_i] = T^{(j)}_i$ (see Lemma \ref{lem:CS} in Appendix).  This expectation is only over randomness in the choice of $h \in \mathcal{H}$ and $s \in \mathcal{S}$.  

Next we consider how large a discord score of time series dimension $T^{(j)}$ needs to be to show up as the discord in the sketched time series $\textbf{R}^{(g)}$
We provide a fairly crude bound that does not depend on the randomness of the data, only using that the dimensions are each $z$-normalized.  Set $k = \sqrt{d}$ and consider a discord $T^{(j)}_{i,m}$ that has distance of $\|\Delta\| = dist(T^{(j)}_{i,m},T^{(j)}_{i',m})$ where $T^{(j)}_{i',m}$ is the nearest neighbor of $T^{(j)}_{i,m}$ in the training set.  Then if $\|\Delta\| > \tau = \frac{1}{\sqrt{\delta}} m d^{1/4}$, then it will be detected as a discord in $\textbf{R}^{(g)}$ with probability at least $1-\delta$.  

The above bound requires $\tau$ grows with $d^{1/4}$ and generally must be much larger than $1$ standard deviation in data value, so we analyze a stronger bound if we assume some noisy periodic structure in each time series.  Under the \emph{$\eta$-periodic assumption}, we assume each time series basically has a repeated structure where each element deviates from that sequence with standard deviation at most $\eta$ (after $z$-normalizing).  
Under this setting, with high probability (the probability of failure decreases exponentially as $n_{\text{train}}$ grows), if $\|\Delta\| > \tau > 2 m \eta$, then it will be detected as the discord in $\textbf{R}^{(g)}$ (see Lemma \ref{lem:noise} in Appendix).

\section{Experiments}
\label{sec:exp}
We first employ synthetic data to assess the quality of the approximated solution. 
Next, we provide case studies with a public transportation dataset and a financial transactions dataset to demonstrate the broad applicability of the proposed method.
Finally, we utilize a water treatment/distribution system dataset to evaluate the effectiveness of the proposed algorithm in the anomaly detection task. 
\textit{It is important to note that anomalies and discord are distinct concepts.}
Anomalies are defined within the specific context and semantics of the dataset, often annotated manually. 
Conversely, discords are domain-independent, can be used to identify potential anomalies, and are explicitly defined using Def.~\ref{def:multi_discord}.
The source code can be found in~\cite{supplementary}.

\subsection{Synthetic Data}
\label{sec:exp_synth}
In this section, we use synthetic data to evaluate the quality of the approximated solution compared to the exact solution.
The quality of the solution is measured by how well the proposed algorithm solves the equation from Def.~\ref{def:multi_discord}.

\noindent \textbf{Experiment Setup:}
We generate synthetic datasets by creating random walk time series. 
It is important to note that random walk time series pose a significant challenge for time series discord mining, as the discord does not exhibit a visually distinct pattern compared to the rest of the time series.
The length of the generated time series is fixed at 10,000, and we vary the dimensionality $d$ of the time series. 
We specifically focus on exploring different values of $d$ since it is the primary challenge addressed by our proposed method. 
We set $d$ to the following values: 250, 500, 1,000, 2,500, 5,000, 7,500, and 10,000.
During the experiment, we use a subsequence length of 100. 
The experiment is repeated 100 times. 
We set hyperparameter $k$ to $\lceil \sqrt{d} \rceil$ to optimize the $O(k + d/k)$ term from combining phases of detection Algs. \ref{alg:detection} and \ref{alg:dim-detection}.  

\noindent \textbf{Performance Measurement:}
The objective of this section is to evaluate the accuracy of the approximated solution in comparison to the exact solution.
To begin with, we calculate the discord score, which represents the nearest neighbor distance, for each subsequence across all dimensions.
Subsequently, we generate a ranked list by assigning ranks to all the subsequences based on their discord scores. 
To measure the quality of the approximated solution, we compute the success rate over 100 trials.
We consider the approximated algorithm successful when the identified discord is ranked within the top 0.01\% of the list.

\begin{figure}[htbp]
\centerline{
\includegraphics[width=0.95\linewidth]{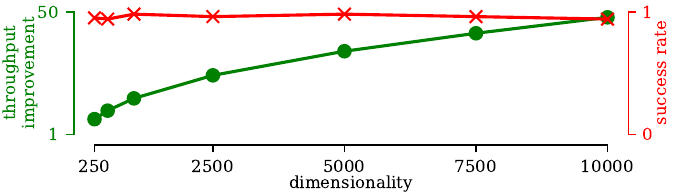}
}
\caption{
The proposed method improves throughput by 50X while maintaining almost perfect success rate when d = $10000$.
}
\label{fig:scale_d}
\end{figure}

\noindent \textbf{Result:}
The results of the experiments are illustrated in Fig.~\ref{fig:scale_d}. 
The speedup presented takes into account the total runtime, which includes both the sketching and detection processes.

The proposed method demonstrates a noticeable improvement in terms of speedup as we increase the value of $d$.
At the highest tested value of $d$ (i.e., 10,000), the speedup reaches 50X with an almost 100\% success rate.

\begin{figure}[htbp]
\centerline{
\includegraphics[width=0.9\linewidth]{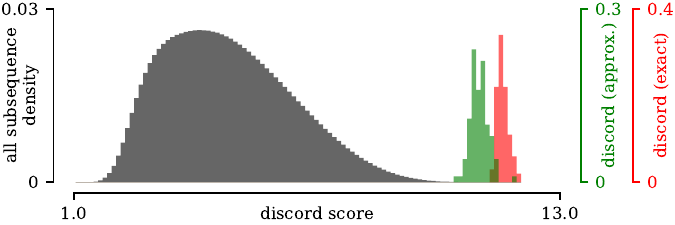}
}
\caption{
The density distribution of discord scores for all subsequences (\textcolor{gray}{gray}), the discords found using the proposed approximation method (\textcolor{green}{green}), and the discords found using the exact method (\textcolor{red}{red}).
}
\label{fig:score_dist}
\end{figure}

To gain further insights into the quality of the approximate solution, we analyze the distribution of discord scores (i.e., the nearest neighbor distance) for three sets of subsequences.
The first set encompasses all subsequences, representing the overall distribution of discord scores. 
The second set comprises the true discords discovered using the exact algorithm, while the last set consists of the discords identified using the proposed sketching method.
Fig.~\ref{fig:score_dist} displays the density plots illustrating the distribution associated with each set of subsequences.
The $d$ is set to 1,000 in this figure.
Notably, the distribution of discords obtained through the proposed approximate method is distinctly different compared to the distribution of discord scores for all subsequences. 
The distribution of the approximated discords is much more similar to the distribution of discords obtained through the exact method.
It is important to note that the task of discord mining in the random walk dataset presents a greater level of difficulty compared to other datasets. 
This increased challenge arises from the fact that the discord, representing the most unusual subsequence (red distribution), is not as easily distinguishable from the other subsequences.
In the real datasets (see Section~\ref{sec:exp_water} and Section~\ref{sec:exp_label}), the discord scores between the exact discord and the sketched ones are much smaller.
In Fig.\ref{fig:mrt_dist}, the discord score of the subsequence discovered by the exact algorithm deviates from the mean of the subsequences discovered by the approximated algorithm by 1.97 standard deviations. 
Similarly, in Fig.\ref{fig:vdata_dist}, the discord score of the subsequence discovered by the exact algorithm deviates from the mean by 2.11 standard deviations.

\subsection{Taipei Mass Rapid Transit (MRT)}
\label{sec:exp_mrt}
The Taipei MRT system is a metro system that moves over two million people daily in Taipei and its surrounding satellite towns.
It could be actionable to bring the operator's attention to the unusual patterns in the ridership data.
For example, a traffic manager may spot an anomaly, perhaps a sudden exodus from a station caused by customers fleeing a crime~\cite{huang2014critically}, and she may be able to intervene stopping trains before they arrive at that station. 
In this section, we apply the proposed algorithm to discover multidimensional time series discord from the per-hour ridership data.
Discords typically correspond to interesting and unique events.

\noindent \textbf{Dataset:}
The original dataset is downloaded from~\cite{datataipei}, and we use the version organized by the authors of~\cite{yeh2019online}.
The dataset consists of time series measuring the number of people entering and exiting each station per hour from November 1, 2015 to March 31, 2017.
There are 108 stations in the dataset.

\noindent \textbf{Experiment Setup:}
We focus on the time series corresponding to the number of people entering and exiting each of the 108 stations.
We set the subsequence length~$m$ to two days (i.e., 48) and the sketching parameters to $k=\lceil \sqrt{d} \rceil$ when applying the proposed method.
A list of discovered discords is returned by the algorithm order based on the nearest neighbor distance (i.e., discord score) from large to small.
We ignore citywide incidents like typhoon days when inspecting the returned discords.

\begin{figure}[htbp]
\centerline{
\includegraphics[width=0.95\linewidth]{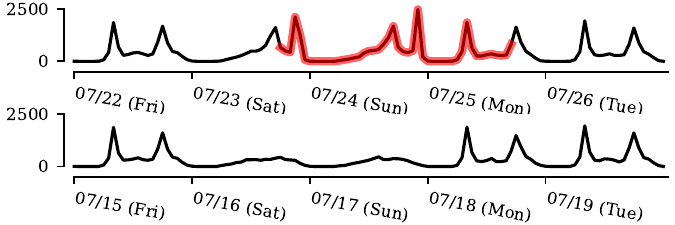}
}
\caption{
The time series is from Nangang Software Park station in 2017.
The top figure shows the discovered time series discord (\textcolor{red}{red}) with additional context.
The $y$-axis is the number of people entering and exiting the station per hour.
}
\label{fig:nongang_soft}
\end{figure}

\noindent \textbf{Result:}
The discord with the largest discord score is shown in Fig.~\ref{fig:nongang_soft}, and it is discovered from \textit{Nangang Software Park Station}.

In the top figure, we show the discovered discord with extra dates preceding and following the discord for context.
Because the area around Nangang Software Park Station consists mostly of office buildings, it has morning and afternoon peaks during workdays.
The afternoon peaks that happen during the weekend are very unusual for the station.
As shown in Fig.~\ref{fig:nongang_soft}.\textit{bottom}, the time series from the previous week has no peak during the weekends.
In other words, the discovered discord indeed captures some special events that happen around the station.
After inspecting the area surrounding the station on a street map, there is a parking lot called \textit{Nangang C3 Field}~\cite{nangangc3} that can also be used as a concert ground for 40,000 attendances.
Upon inspecting the event schedule for the venue~\cite{nangangc3}, there were indeed concerts held on both 7/23 and 7/24 by a popular rock band Mayday~\cite{maydayband}.
The rock concert explains the peaks/discord discovered on 7/23 and 7/24.

\begin{figure}[htbp]
\centerline{
\includegraphics[width=0.9\linewidth]{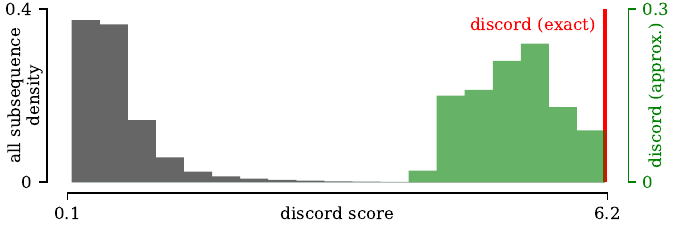}
}
\caption{
The density distribution of discord scores for all subsequences (\textcolor{gray}{gray}) and the discords found using the approximation method (\textcolor{green}{green}).
The discord score of the discord found using the exact method is indicated by a red line.
}
\label{fig:mrt_dist}
\end{figure}

In addition to the exploratory analysis, we compare the distribution densities using a methodology similar to that described in Section~\ref{sec:exp_synth}.
The resulting figure is presented in Fig.~\ref{fig:mrt_dist}.
It is evident that the discovered discords are clearly outliers when compared to the distribution of all subsequences.
Furthermore, the proposed method achieved an average speedup of 3.6X compared to the exact solution.
Please note that in this particular case, there is only one dataset, and as a result, there is also only one true discord indicated by a single line.

\subsection{Payment Network.}
\label{sec:exp_label}
There are millions of transactions among different entities processed daily on a modern payment network.
It is important to monitor the activity of these entities for unusual patterns (i.e., discords) as such events could have undesirable effects on the software or hardware infrastructures built around the payment network.
For example, a sudden fluctuation in the transaction volume could lead to undesirable declines if the fluctuation is not attended to.
To demonstrate the utility of the proposed method, we apply our discord mining algorithm to the per-hour transaction volume multidimensional time series where different dimensions are different categories of merchants.
Similar to Section~\ref{sec:exp_mrt}, multiple discords are returned by our algorithm and we order them based on the ``discordness" of the returned pattern (i.e., the distance with the pattern's nearest neighbor).

\noindent \textbf{Dataset:}
We prepare the multidimensional time series from our internal transaction database.
The per category per hour time series is aggregated using transaction data from January 1st, 2018 to July 1st, 2021.
The length of the time series is 30,600, and the number of categories is around 1,000.

\noindent \textbf{Experiment Setup:}
When applying the proposed method, we set the subsequence length~$m$ to 36 or 1.5 days.
The sketching parameter is set to $k=\lceil \sqrt{d} \rceil$.
Because it is beneficial to examine the second/third discord in the time series, our algorithm once again returns multiple discords as we did in Section~\ref{sec:exp_mrt}.
The discords are ordered based on the nearest neighbor distance (i.e., discord score) from large to small.

\begin{figure}[ht]
\centerline{
\includegraphics[width=0.9\linewidth]{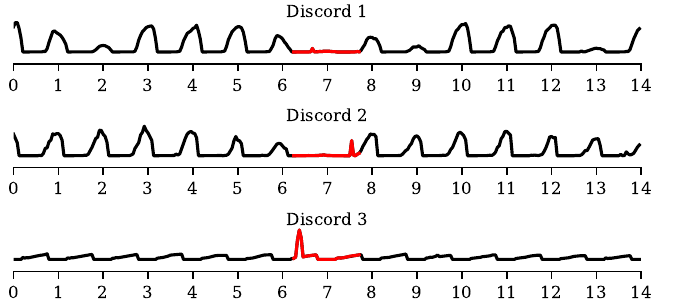}
}
\caption{
The top three discovered time series discords.
The discords (\textcolor{red}{red}) are plotted with additional context, and total of 14 days (336 hours) of time series are plotted.
}
\label{fig:vdata}
\end{figure}

\noindent \textbf{Result:}
The top three discovered discords are shown in Fig.~\ref{fig:vdata}.

The first discovered discord (i.e., Discord 1) consists of a period of time with relatively small numbers of transactions with a small bump around the 7th-day mark.
The second discovered discord (i.e., Discord 2) also captures a period of time where the number of transactions is relatively low.
However, there is a sharp spike near the middle of the 7th day.
One possible reason for the low transaction volume could be that majority of merchants from the corresponding categories are closed during that time.
The small bump around the 7th-day mark in Discord 1 indicates there might be a subset of merchants still doing business during that time.
For Discord 2, the sharp spike that happens right before the normal business volume around the 8th-day mark could mean that merchants of the Discard 2 category could still be accepting orders, but push the payment process to the beginning of the next business day.
The third discovered discord (i.e., Discord 3) could indicate the occurrence of a special sales event.
Time series discords are capable of capturing both high and low-volume events that occur in the payment network.
The speedup of the proposed method in throughput relative to the exact solution is 13 for this dataset.

\begin{figure}[htbp]
\centerline{
\includegraphics[width=0.9\linewidth]{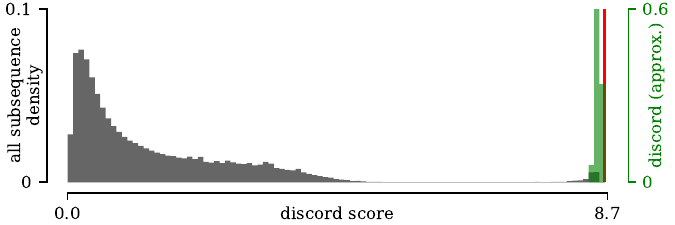}
}
\caption{
The density distribution of discord scores for all subsequences (\textcolor{gray}{gray}) and the discords found using the approximation method (\textcolor{green}{green}).
The discord score of the discord found using the exact method is indicated by a red line.
}
\label{fig:vdata_dist}
\end{figure}

We further compared the distribution densities similar to the experiment conducted with the MRT data.
The resulting figure is displayed in Fig.~\ref{fig:vdata_dist}, showcasing that the discovered discords are distinctly outliers in comparison to the distribution of all subsequences. The discord identified by the approximate method exhibits a significantly higher similarity to the exact discord compared to the synthetic dataset (Figure \ref{fig:score_dist}), which demonstrates the effectiveness of the sketching algorithm on the real-world dataset.

\subsection{Water Treatment and Distribution}
\label{sec:exp_water}
In order to showcase the efficacy of the approximate sketching algorithm in anomaly detection, we employ a dataset obtained from a water treatment and distribution system. Our primary objective is to conduct a quantitative comparison between the proposed discord mining algorithm and other existing methods commonly used for anomaly detection.

\noindent \textbf{Dataset:}
We performed experiments with SWaT~\cite{goh2016dataset} and WADI datasets~\cite{ahmed2017wadi} in detecting attacks on cyber physical systems.
SWaT dataset contains the measurement of a scaled-down version of a real-world industrial water treatment plant under the normal and attacked scenarios~\cite{goh2016dataset}.
A total of 51 sensors and actuators are measured for 11 days with a 1 Hz sampling rate.
The first 7 days consist of the measurement from normal operation. 
The system is attacked 41 times during the last 4 days.
WADI dataset consists of measurements from a water distribution system with 123 sensors and actuators~\cite{ahmed2017wadi}.
The measurement takes place in 16 days where the system operates normally in the first 14 days and is attacked 14 times in the last 2 days.
The data sampling rate is 1 Hz.

\noindent \textbf{Experiment Setup:}
We compare the proposed approximated multidimensional discord mining matrix profile (Discord/Fast) to methods like exact multidimensional discord mining matrix profile (Discord/Exact), 1NN~\cite{pedregosa2011scikit}, LOF~\cite{breunig2000lof,pedregosa2011scikit}, OC-SVM~\cite{scholkopf2001estimating,chang2011libsvm,pedregosa2011scikit}, and MAD-GAN~\cite{li2019mad}.
MAD-GAN is one of the state-of-the-arts for the datasets, and the other methods are classic anomaly detection algorithms~\cite{li2019mad}.

We follow the following procedure to apply discord-based methods (Discord/Fast and Discord/Exact) to the anomaly detection problem.
First, we identify the dimension that contains the multidimensional discord using either the exact multidimensional discord mining algorithm or the proposed sketching algorithm. 
Let us assume that the discord is found in dimension $j$. In the anomaly detection test dataset, we have the label of each subsequence, so the next step is to compute the anomaly score associated with each subsequence. We extract the $j$th dimension of the test time series $T^{(j)}_{\text{test}}$ and join it with $T^{(j)}_{\text{train}}$ using the matrix profile algorithm.
In other words, for each subsequence in $T^{(j)}_{\text{test}}$, we find its nearest neighbor in $T^{(j)}_{\text{train}}$ and compute the distance between each pair of nearest neighbors. 
We use these nearest neighbor distances as the anomaly score for each subsequence in $T^{(j)}_{\text{test}}$.
We set $k=\lceil \sqrt{d} \rceil$ when using the Discord/Fast method.


\noindent \textbf{Performance Measurement:}
With the label and anomaly score of each subsequence in $T^{(j)}_{\text{test}}$, the performance measurements adopted for both datasets are ROC-AUC (AUC) scores.
We choose to use AUC score instead of the more common F1 score for these datasets because the anomaly scores need to be converted to binary predictions for evaluating the F1 score of a method.
As pointed out by Kim et al.~\cite{kim2021towards}, converting the anomaly scores to binary labels for time series anomaly detection is itself a challenging problem.
Since our paper is focusing on mining time series discord, we use AUC score to avoid the additional complication associated with the anomaly score conversion.

\begin{table}[ht]
\centering
\caption{
The proposed method achieves comparable performance to the best alternative with reduced runtime.}
\label{tab:water}
\footnotesize
\begin{tabular}{l||cc|cc}
 & \multicolumn{2}{c|}{SWaT} & \multicolumn{2}{c}{WADI} \\
Method & AUC & Time (sec.) & AUC & Time (sec.) \\ \hline \hline
1NN & 0.82 & 8,534 & 0.47 & 5,057 \\ 
LOF & 0.79 & 18,778 & 0.47 & 30,241 \\
OC-SVM & 0.82 & 42,934 & 0.52 & 130,221 \\
MAD-GAN & 0.81 & 3,273 & 0.45 & 3,627 \\ \hline
Discord/Exact & \textbf{0.83} & 2,488 & 0.62 & 2,544 \\
Discord/Fast & 0.76 & \textbf{821} & \textbf{0.72} & \textbf{653}
\end{tabular}

\end{table}

\noindent \textbf{Result:}
The AUC scores for all the tested methods on both datasets are summarized in Table~\ref{tab:water}.
For the SWaT dataset, all methods achieve comparable AUC scores.
Since the proposed method also has similar performance in terms of AUC, its advantage in running time makes it more desirable than the others.
The performance difference between different methods on the WADI dataset is even greater.
The discord-based methods are noticeably better than the alternatives.
Note that the dimension located by the exact discord-based method could sometimes be inferior compared to the one found by approximated method because accurate discord mining does not necessarily translate to accurate anomaly detection. 

To evaluate the robustness of different methods, we add 200 random walk time series to each dataset and redo the experiments in Table~\ref{tab:water_add}. 
The expanded SWaT dataset consists of 251 and the expanded WADI consists of 323 dimensions. 

\begin{table}[ht]
\centering
\caption{
The proposed method is robust against added random walk dimensions.}
\label{tab:water_add}
\footnotesize
\begin{tabular}{l||cc|cc}
 & \multicolumn{2}{c|}{SWaT} & \multicolumn{2}{c}{WADI} \\
Method & AUC & Time (sec.) & AUC & Time (sec.) \\ \hline \hline
1NN & 0.66 & 9,857 & 0.47 & 6,010 \\
LOF & 0.76 & 28,282 & 0.42 & 35,038 \\
OC-SVM & 0.79 & 186,623 & 0.47 & 368,884 \\
MAD-GAN & 0.49 & 7,814 & 0.46 & 5,318 \\ \hline
Discord/Exact & \textbf{0.83} & 12,755 & 0.62 & 7,731 \\
Discord/Fast & 0.76 & \textbf{2,261} & \textbf{0.64} & \textbf{1,382}
\end{tabular}
\end{table}

When contrasting the performances in Table~\ref{tab:water_add} with the ones in Table~\ref{tab:water}, the SWaT dataset's AUC degrades more compared to the WADI dataset.
Out of all the tested methods, LOF and the proposed method are more robust against the added noise.
For the discord-based methods, we found that the discovered discords are from one of the original dimensions rather than the added noisy dimensions.
The proposed discord-based methods better ignore the noisy dimensions when detecting an anomaly.
The speedup of the approximated method compared to the exact method is around 5.6X for both datasets.
\section{Conclusion}
The paper proposed an efficient sketching-based multidimensional discord mining algorithm to quickly identify the discord.
The algorithm improves the throughput by almost 50x and provably maintains high accuracy compared to the exact discord mining algorithm.
Experiments are performed with four datasets from various domains to demonstrate the effectiveness and utility of the proposed algorithm.
For future work, we consider extending the sketching-based method for subsequence classification~\cite{yeh2023egonetwork} or to explore the data privacy issues associated with time series~\cite{yeh2023time}.

\section*{Acknowledgement}
Jeff M. Phillips thanks NSF III-1816149.

\Urlmuskip=0mu plus 1mu\relax
\bibliographystyle{IEEEtran}
\bibliography{section/ref}

\appendix

\vspace{-1mm}
In this section we provide an analysis of why our algorithm provides high probability of recovery.  We borrow from the analysis of the classic Count-Sketch~\cite{charikar2002finding} for finding heavy hitters among data streams  from a domain size $[d]$.

\paragraph{Modeling of Algorithm.}
As mentioned above, we say $h$ is drawn from a 2-universal family $\mathcal{H}$ of hashing functions, that is, so that for any two distinct dimensions $j, j' \in [d]$ that $\textsf{Pr}_{h \sim \mathcal{H}}[h(j) = h(j')] = 1/k$.  


We now argue that when we estimate the value $T^{(j)}_i$ of the $j$th time series at a time point $i$.  In particular, we use $s(j) R^{(g)}_i$ where $j \in J_g$ so $g = h(j)$.  Following the standard analysis of the Count-Sketch~\cite{charikar2002finding} we obtain the following.  

\begin{lemma_}\label{lem:CS}
For $g = h(j)$ then for any time point $i$ we have 
\\ an unbiased estimate $E[s(j)R^{(g)}_i] = T^{(j)}_i$ 
\\ and its variance is 
$Var[s(j)R^{(g)}_i] = \frac{1}{k}\sum_{j' \neq j} T^{(j')}_i$.  
\end{lemma_}
\begin{proof}
Let $Y_{j',g} = 1$ if $h(j') = g$ and $0$ otherwise.  
The expected value of the sketched time series $s(j) R^{(g)}_i$ for $g = h(j)$ can be factored as follows:
{\footnotesize\[ 
E[s(j) R^{(g)}_i] = s(j) s(j) T^{(j)}_i + \sum_{j' \neq j} E[Y_{j',g} s(j) s(j') T^{(j')}_i].
\]}
Since $s$ is pairwise independent and independent of $h$
{\footnotesize
\begin{align*}
E[Y_{j',g} s(j) s(j') T^{(j')}_i] 
&= 
E[Y_{j',g} s(j)] E[s(j')] T^{(j')}_i 
\\ &= 
E[Y_{j',g} s(j)] \cdot 0 \cdot T^{(j')}_i = 0.
\end{align*}}
All that remains is $s(j) s(j) T^{(j)}_i = T^{(j)}_i$, as desired.  

To bound the variance, using our expectation bound, we calculate
{\footnotesize
\begin{align*}
Var[s(j) R^{(g)}_i] 
 &= 
E[(s(j) R^{(g)}_i - E[s(j) R^{(g)}_i])^2] 
 \\ &= 
E[(\sum_{j' \neq j} s(j) s(j') Y_{j',g} T^{(j')}_i)^2]
 \\ &=
E[\sum_{j' \neq j} \sum_{j'' \neq j} s(j'') s(j') Y_{j',g} Y_{j'',g} T^{(j')}_i T^{(j'')}_i]
 \\ &=
\sum_{j' \neq j} \sum_{j'' \neq j} T^{(j')}_i T^{(j'')}_i E[s(j'') s(j') Y_{j',g} Y_{j'',g}]
\end{align*}}
Since $s$ is pairwise independent, $E[s(j') s(j'')] = E[s(j')]E[s(j'')] = 0$ for $j' \neq j''$.  Hence all that remains from the variance are terms when $j' = j''$ as
{\footnotesize\begin{align*}
Var[s(j) R^{(g)}_i] 
  &= 
\sum_{j' \neq j} (T^{(j')}_i)^2 E[Y_{j',g}^2]
 = 
\sum_{j' \neq j} (T^{(j')}_i)^2 E[Y_{j',g}]
 \\ &= 
\frac{1}{k} \sum_{j' \neq j} (T^{(j')}_i)^2. \qedhere
\end{align*}}
\end{proof}

Now recall that we have $z$-normalized every time series, so the expected value of $(T^{(j')}_i)^2$ (over the choice of $t$) is $1$.  Thus
{\footnotesize\[
Var[s[j] R^{(g)}_i] = \frac{1}{k} \sum_{j' \neq j} (T^{(j')}_i)^2 = \frac{d-1}{k}.
\]}

\paragraph{Analysis of Subsequences with Large Discord Scores.}
Next consider that time series $T^{(j)}$ has a discord with large discord scores, we can now show it is likely to appear as a discord in $R^{(g)}$ -- the sketched time series of $\textbf{T}$, where $g = h(j)$.  
Considering the time series with subsequences of length $m$, for a subsequence $T_{i,m}$ in the test set, if the closest subsequence in the training set $T^{(j)}_{i',m}$ is a perfect match (e.g., $T^{(j)}_{i,m} = T^{(j)}_{i',m}$), then we have that for the corresponding spots in the associated sketched time series $R^{(g)}$ that $E[s(j) R^{(g)}_{i,m}] = T^{(j)}_{i,m} = T^{(j)}_{i',m} = E[s(j) R^{(g)}_{i',m}]$.  
On the other hand, consider a discord that has large discord score, so $T_{i,m} - T_{i',m} = \Delta \in \mathbb{R}^m$, and $\|\Delta\|$ is large.  In this case 
{\footnotesize
\[
E[s(j) R^{(g)}_{i,m}] = T^{(j)}_{i,m} = T^{(j)}_{i',m} + \Delta = E[s(j) R^{(g)}_{i',m}] + \Delta.
\]}
So in expectation, the large discord score $\Delta$, measured on the aggregated times series is in expectation still $\Delta$.  

We want to understand how large does $\|\Delta\|$ need to be to show up against the noise inherent in the sketched aggregation step.  To do this, we can calculate the variance of a subsequence.  In particular we have 
{\footnotesize\begin{align*}
Var[R^{(g)}_{i,m}] 
&= 
Var[ \sum_{t \in [i,...,i+m]} R^{(g)}_t] 
= 
m^2 Var[R^{(g)}_{i,m}] = m^2 \frac{d-1}{k}.
\end{align*}}

So if $\|\Delta\|^2 = \|R^{(g)}_{i,m} - R^{(g)}_{i',m}\|^2 = dist(R^{(g)}_{i,m},R^{(g)}_{i',m})^2$ is much larger than the variance, it will be consistently detected.  We can formalize this with a Chebyshev bound.  
{\footnotesize\[
Pr[ dist(R^{(g)}_{i,m} , R^{(g)}_{i',m})^2 \geq \alpha] 
\leq 
\frac{Var[R^{(g)}_{i,m}]}{\alpha^2} 
= 
\frac{m^2 \frac{d-1}{k}}{\alpha^2}.
\]}
If we set $k = \sqrt{d}$ (as we do in the implementation), and $\alpha = \frac{1}{\sqrt{\delta}} m d^{1/4}$, then we have for any $\delta \in (0,1)$ that
{\footnotesize\[
Pr[ dist(R^{(g)}_{i,m} , R^{(g)}_{i',m}) \geq \frac{1}{\sqrt{\delta}} m d^{1/4}] \leq \delta.  
\]}
Thus if the discord is sufficiently large, it will be detected with high (at least $1-\delta$) probability.  
Notably, this is with regard to the randomness in the selection of the hash functions (placement in group, and  the \{-1,+1\} choice), and not randomness in the data.  So this bound is assumption free and can handle adversarial data.

\paragraph{Modeling of Non-adversarial Data.}
We can refine this analysis to show high probability of recovering a large discord by considering non-adversarial modeling of the data.  
Without these assumptions, adversarial unlikely settings may occur. For instance, if each test data subsequence has exactly one similar subsequence in the training data (standard discord analysis will show small discord scores, but it is not robust).  Moreover, these matches could be misaligned temporally across dimensions, so the sketched time series presents these non-robust matches as significant discords.  If many such cases occur, this can causes a large difference between the discord found by searching sketched time series (as in our algorithm) and from searching the time series for discords individually.  

Specifically we now consider an \emph{$\eta$-periodic assumption}:  that is, that each time series $T^{(j)}$ has a period $p_j$ (it has a rough pattern that repeats after $p_j$ time steps), and for any two time series $T^{(j)}$ and $T^{(j')}$ that $p_j$ and $p_{j'}$ have a small common factor.  That is, there exist small positive integers $a_j$ and $a_{j'}$ so that $P = p_j a_j = p_{j'} a_{j'}$.  This ensures that all time series have some common latent structure which controls their frequency, and every period $P$ they re-align.  For example, periods could be daily, weekly, or hourly and moreover common structural shifts (e.g., day-light-savings-time) affect the latent structure but keeps the general periodic alignment in place.  
This assumption ensures that for each time window in the test data, that for each period $P$ in the test data there exists a corresponding time window in the training data that has a similar phase across \emph{all} individual time series. 
Within each alignment there may of course be small fluctuations, and noise, but not structural ones -- unless there is a true large discord.  In particular, we assume after $z$-normalizing, that each $T^{(j)}_i$ has variance $Var[T^{(j)}_i] \leq \eta^2$.  So if $T^{(j)}_i$ in the test aligns to its corresponding point $T^{(j)}_{i'}$ in the training, their difference has expected value $E[T^{(j)}_i - T^{(j)}_{i'}] = 0$ and variance $Var[T^{(j)}_i - T^{(j)}_{i'}] \leq 2\eta^2$. 
This is interesting when $\eta$ is significantly smaller than $1$.  
And we assume there are many periods of length $P$ within the span of the training data, and thus the difference in the number of periods for an individual time series of length $p_j$ will not be too different from the number of periods of length $P$. 


Now analyzing the variance under the periodic assumption, we show with high probability that if there is a large discord score, the sketched time series preserves it.  

\begin{lemma_}\label{lem:noise}
Under the $\eta$-periodic assumption with joint period $P$, then if there is a single subsequence $T^{(j)}_{i,m}$ of size $m$, with most similar subsequence in the training data as $T^{(j)}_{i',m}$ and $dist(T^{(j)}_{i,m},T^{(j)}_{i',m}) > 2 \eta m$, then with high probability (at least $1-d n_{\text{test}}/2^{n_{\text{train}}/P})$, that subsequence will be detected.  
\end{lemma_}
\begin{proof}
Via Lemma \ref{lem:CS} we have $Var[s(j)R^{(g)}_i] \leq \frac{1}{k} \sum_{j' \neq j} T^{(j')}_i$, where $g = h(j)$.   
Now on a subsequence from the $j$th time series, $T^{(j)}_{i,m}$ in the test data, and its periodic match in the training data $T^{(j)}_{i',m}$, we can analyze their difference $\Delta = T^{(j)}_{i,m} - T^{(j)}_{i',m}$.  By the $\eta$-periodic assumption that assumes for aligning elements $E[T^{(j)}_{i,m}(t) - T^{(j)}_{i',m}(t)]=0$ then and $E[\Delta(t)] = 0$ for each index $t$.  Moreover, using that $Var[T^{(j)}_{i,m}(t)] \leq \eta^2$ we have 
{\footnotesize
 \begin{align*}
 E[\|\Delta\|^2] 
 &= 
 \sum_{t = 1}^m E[(T^{(j)}_{i,m}(t) - T^{(j)}_{i',m}(t))^2] 
 \\ &=   
 \sum_{t = 1}^m E[(T^{(j)}_{i,m}(t) - T^{(j)}_{i',m}(t))^2- E[T^{(j)}_{i,m}(t) - T^{(j)}_{i',m}(t)]^2] 
 \\ &=   
 \sum_{t = 1}^m Var[(T^{(j)}_{i,m}(t) - T^{(j)}_{i',m}(t))]
 \\ & \leq
 \sum_{t = 1}^m Var[T^{(j)}_{i,m}(t)] + Var[T^{(j)}_{i',m}(t)]
 \leq 
 2m\eta^2. 
 \end{align*}}
Now we bound the variance of $\|\Delta\|^2$ assuming that noise on each time point is independent.  
{\footnotesize\begin{align*}
Var[\|\Delta\|^2] 
 &= 
E[\|\Delta\|^4] - E[\|\Delta\|^2]^2
 \\ & \leq 
E[\|\Delta\|^4]
 = 
\sum_{t, t' \in [1...m]} E[\Delta(t)^2\Delta(t')^2]
 \\ &= 
 \sum_{t, t' \in [1...m]} E[\Delta(t)^2] E[\Delta(t')^2]
 \\ &= 
 \sum_{t, t' \in [1...m]} (2m \eta^2) (2m \eta^2)
 = 
 4m^4 \eta^4 
\end{align*}}
Recall that a Chebyshev inequality for a random variable $X$ states that $Pr[|X - E[X]| \geq \alpha ] \leq Var[X]/\alpha^2$ for any $\alpha > 0$.   
Now assume there is a discord with large discord scores that has $dist(T^{(j)}_{i,m},T^{(j)}_{i',m}) = \tau$, we can bound the probability at another in-period sequence pair $T^{(j)}_{q,m},T^{(j)}_{q',m}$ with $\Delta' = T^{(j)}_{q,m} - T^{(j)}_{q',m}$ exceeds that distance with a Chebyshev bound
{\footnotesize
\begin{align*}
Pr[dist(T^{(j)}_{q,m},T^{(j)}_{q',m}) > \tau]  
&\leq
Pr[ |\|\Delta'\|^2 - E[\|\Delta'\|^2]|  >  \tau^2 \hspace{-1mm}- \hspace{-1mm}E[\|\Delta'\|^2]] 
\\&\leq 
\frac{Var[\|\Delta'\|^2]}{(\tau^2 - E[\|\Delta'\|^2])^2}.  
\end{align*}}
If we set $\tau^2 = 4m \eta^2(1 + m/\sqrt{\delta})$ or more simply $\tau > 2m\eta \delta^{-1/4}$, then with some algebra we can achieve 
{\footnotesize\[
Pr[dist(T^{(j)}_{q,m},T_{q',m}) > \tau] \leq \delta.
\]}

This is for a simple potential match.  If there are $n' = n_{\text{train}}/P$ potential matches (under the periodic assumption), then the distance must exceed this for all $n'$ matches.  Let $\|\Delta'_*\| = \min_{\ell \in 1 \ldots n'} dist(T^{(j)}_{q,m},T^{(j)}_{q',m})$.  By a union bound, we have $\|\Delta'_*\| > \tau$ only if no $n'$ potential matches exceeds $\tau$ which happens with probability at most $\delta^{n'}$ if their noise is independent.  Setting $\delta = 1/2$, then if $\tau > (2^{3/4})m\eta > 2m\eta$ then
$
Pr[\|\Delta'_*\| \geq \tau] \leq 1/2^{n'}.
$

Next we need to consider the $d n_{\text{test}}$ different subsequences from the $d$ time series which may show up as false discords.  Within a time series, these overlap, so are not independent.  But we can use a union bound to address the concern than if any of these has more than $\tau$ distance from its closest match in the training set it will show up as the discord.  If we set $\delta' = (1/2^{n'}) d n_{\text{test}}$, and have $\tau > 2m\eta$, then with probability at least $1-(1/2^{n'} \cdot d \cdot n_{\text{test}})$, if an discord has $\|\Delta\| > \tau$ it will show up as the discord.  
\end{proof}

\end{document}